\documentclass[fp,twocolumn]{jpsj3}
\usepackage{txfonts}
\usepackage{here}
\usepackage{txfonts} %Please comment out this line unless the txfonts package is availabe in your LaTeX system.

\usepackage{bm}
\usepackage{amsmath}
\usepackage{braket}

\usepackage{color}
\usepackage{subfloat}

\newcommand{\f}{\frac}

\newcommand{\ome}{\omega}

\newcommand{\del}{\partial}

\newcommand{\nab}{\nabla}

\newcommand{\al}{\alpha}

\title{Faraday Waves in Bose--Einstein Condensates\\
$\sim$The Excitation by the Modulation of the Interaction and  the Potential$\sim$}

\author{Nobuyuki Shukuno$^1$\thanks{m21sa015@st.osaka-cu.ac.jp}, Yuto Sano$^1$, and Makoto Tsubota$^{1,2}$}
\inst{$^1$Department of Physics, Osaka City University, Osaka 558-8585, Japan \\
$^2$Department of Physics, Osaka Metropolitan University, Osaka 558-8585, Japan \\} %\\

\abst{We numerically study the dynamics of Faraday waves for Bose--Einstein condensates(BECs) trapped by anisotropic potentials using the three-dimensional Gross--Pitaevskii equation. In previous studies, Faraday waves were excited by periodic modulation of the interaction or potential; in contrast, this study systematically addresses the excitations of the two methods. When the interaction is modulated with a modulation frequency resonant with Faraday waves, the breathing mode along the tight confinement direction is excited, and the Faraday waves appear in the direction of weak confinement. A modulation frequency that is not resonant with Faraday waves does not excite Faraday waves. Thus, the dynamics depend on modulation frequencies. The behavior of the total energy and its decomposition characterize the dynamics.
The excitation of Faraday waves depends on the anisotropy of the potentials as well; Faraday waves are excited only for elongated BECs. We compare the differences of the dynamics in modulation methods. There are no qualitative differences between the modulation of the interaction and potential. When the interaction and potential are simultaneously modulated, Faraday waves are excited but they do not necessarily work additively. To understand this phenomenon as a dynamical system, we choose a few dynamical variables and follow their trajectory in a phase space. The trajectory characteristics of Faraday waves and the breathing mode show that the methods of modulation are not very relevant; determining the target mode to excite is important.}

\begin{document}
\maketitle

\section{Introduction}

Faraday waves have been studied in view of hydrodynamic instability\cite{chandra}.
In 1831, Faraday analyzed the instability of a free surface of classical fluids in a vertically vibrated vessel\cite{faraday}. Standing wave patterns that were named Faraday waves appeared on the free surface. These waves have been studied theoretically and shown to be related to the instability of the Mathieu equation\cite{Mathieclassical}, which describes parametric resonance phenomena in various science and engineering fields.

Faraday waves have been recently studied as a typical example of hydrodynamic instability in quantum fluids. Instabilities such as the Rayleigh--Taylor and Kelvin--Helmholtz instabilities, which are well known in classical fluids, were studied in atomic Bose--Einstein condensates (BECs)\cite{RTI,KHI}. Faraday waves were also observed in quantum fluids such as superfluid $^4\mathrm{He}$, BECs and Fermi superfluids. In superfluid $^4\mathrm{He}$, they have been excited by a setup similar to that of Faraday's original experiment\cite{experimentHe4}. However, unlike the Faraday waves excited in Faraday's original experiment, Faraday waves in BECs\cite{2,3,Mathieint} and Fermi superfluids\cite{Fermi} are excited as periodic spatial patterns of density. Faraday-like patterns in BECs are reported too.\cite{star1,star2}
The advantages of studying this type of hydrodynamic instability in BECs are its controllability and visibility. Controllability means that the shape of BECs can be controlled by changing trap frequencies and the interaction by using Feshbach resonance.\cite{pethicksmith} Visibility indicates that the density profile of BECs can be experimentally visualized. Theoretically, the various phenomena are quantitatively described by the Gross--Pitaevskii(GP) model for a weakly interacting Bose gas\cite{pethicksmith,bosegas}.

Faraday waves are excited by the periodic modulation of the interaction or potential of BECs trapped by anisotropic potentials. When the interaction is modulated, breathing modes are excited along the direction of tight confinement, followed by Faraday waves along the direction of weak confinement, and the system eventually assumes a granular state, which was confirmed by the simulation of the three-dimensional GP equation and the multiconfigurational
time-dependent Hartree method for bosons (MCTDHB)\cite{2}. When the potential is modulated, these typical dynamics of the excitation of Faraday waves are also observed in experiments BECs\cite{3} and the simulation of the GP equation\cite{4,3}. A Mathieu equation can be derived from the GP equation with the modulation of the interaction or potential.\cite{Mathiepot,Mathieint,symmetry}.

In this work, we numerically study the dynamics of Faraday waves for BECs trapped by anisotropic potentials using the 3D GP equation:
\begin{equation}
    i\hbar \f{\del \psi }{\del t}=-\f{\hbar^2}{2m}\nab^2\psi
    +V(\bm r,t)\psi+g(t)|\psi|^2\psi.\label{GPeq}
\end{equation}
Here, $\psi=\psi(\bm r,t)$ is the macroscopic wavefunction,
\begin{equation}
    V(\bm r,t)=\f{m}{2}(\ome_{xy}^2(t)(x^2+y^2)+\ome_z^2(t)z^2)\label{potential}
\end{equation}
is the anisotropic trapping potential, and $m$ is the mass of the particle. We modulate $g(t),\ome_{xy}(t)$ and $\ome_z(t)$ with a period of modulation frequency. When exciting Faraday waves, efficiently injecting energy into a system is important.
The efficiency of energy injection depends on whether collective modes are resonant or not. Thus, we focus on the time evolution of the energy in the GP model with modulations of $g(t)$ and/or $V(\bm r,t)$. As we are interested in the excitation of Faraday waves, we focus on this and do not follow the dynamics after the Faraday waves are excited.
In the mentioned previous studies\cite{3,2,Mathieint}, Faraday waves were realized by modulating the interaction or potential. We chiefly study the modulation of the interaction; we also deal with the modulation of only the potential and the simultaneous modulation of the interaction and potential.

First, we investigate the conditions for the excitation of Faraday waves by the periodic modulation of the interaction $g(t)$. To excite Faraday waves, we modulate the interaction with a frequency of $2\ome_{xy}$ resonant with the breathing mode along the $xy$ direction. For elongated BECs with $\ome_{xy}\gg\ome_{z}$, a breathing mode is excited in the $xy$ direction of tight confinement, followed by Faraday waves in the $z$ direction of weak confinement. These typical dynamics of Faraday waves are consistent with experimental findings\cite{2}. When Faraday waves are excited, the characteristic density pattern appears and the total energy steadily increases, demonstrating resonance. To investigate these dynamics, we decompose the kinetic energy into the $x,y$ and $z$ directions. The kinetic energy along the $xy$ direction increases with the excitation of the direction’s breathing mode. Then, the kinetic energy along the $z$ direction increases rapidly with the excitation of Faraday waves. 
We investigate the dependence of the excitation of Faraday waves on modulation frequencies. Typically, we investigate the modulation frequency $\sqrt{3}\ome_z$, which corresponds to the breathing mode along the $z$ direction. Then, the breathing mode along this direction is excited; however, patterns such as Faraday waves do not appear. The total energy does not increase as much as in the case of Faraday wave excitation.
We also investigate the dependence of the excitation of Faraday waves on the anisotropy $\ome_{xy}/\ome_z$ of the potentials. Faraday waves are excited only for the elongated BECs.

Second, we compare the differences of the dynamics for different modulation methods. There are no differences when Faraday waves are excited by the modulation of the interaction or potential. When the interaction and potential are simultaneously modulated, Faraday waves are excited. However, both modulations do not necessarily work additively for the excitation of Faraday waves.

Third, we characterize the excitation of Faraday waves by choosing a few dynamical variables. When the total energy is decomposed into kinetic energy $E_{\rm kin}$, potential energy $E_{\rm pot }$, and interaction energy $E_{\rm int}$, like in Eq.\eqref{energy_nondim}, the dynamics of the system can be characterized by a trajectory in 3D $(E_{\rm kin },E_{\rm pot },E_{\rm int})$ space. When Faraday waves are excited, the dynamics follow a characteristic trajectory.

The highlights of this work are as follows. 

\begin{enumerate}
    \item In the previous studies, Faraday waves were excited by
    periodic modulation of either the interaction or potential. This study systematically addresses the excitations by
    the two methods.

    \item We study how the excitation of Faraday waves depends on the modulation frequency and the anisotropy of the potential $\ome_{xy}/\ome_z$. 
    
    \item The story of the dynamics from the breathing mode to Faraday waves is revealed by the decomposition of energy.
    \item The choice of a few dynamical variables casts the dynamics to a simple dynamical system. 
\end{enumerate}

\section{Theoretical model and numerical calculation}

We numerically study the dynamics of Faraday waves for the BECs described by Eqs.\eqref{GPeq} and \eqref{potential}. In our numerical simulation, we consider a gas of $8\times 10^5$ $^7\mathrm{Li}$ atoms. The interaction $g(t)$ is modulated as
\begin{equation}
    g(t)=\f{4\pi\hbar^2}{m}a(t),\,\,\,\,a(t)=a_{\rm bg}\bigg[1-\f{\Delta}{B_0-B(t)}\bigg],
\end{equation}
where $a_{\rm bg}=-24.5a_0$, $a_0$ is the Bohr radius,  $\Delta=192.3{\rm G}$ is the width of the resonance and $B_0=736.8{\rm G}$ is the location of the resonance. These parameters are based on Ref. [8]. The magnetic field was modulated by $B(t)=\bar{B}+\Delta B\sin(\ome_{\rm int} t)$, where $\bar{B}=577.4{\rm G}$, $ \Delta B=5{\rm G} $ is the modulation amplitude, and $\ome_{\rm int}$ is the modulation frequency of the interaction. These parameters are based on Ref. [18]. Thus, $g(t)$ is expressed as
\begin{equation}
    \begin{split}
        g(t)&=\f{4\pi\hbar^2}{m}(-24.5a_0)\bigg(1-\f{192.3}{736.8-(577.4+5\sin\ome_{\rm int} t)}\bigg)\\
        &\approx \f{4\pi\hbar^2}{m}\f{24.5a_0\times159.4}{32.9}\bigg(1+0.04\sin \ome_{\rm int} t\bigg)\\
        &\equiv g(1+0.04\sin \ome_{\rm int} t).
    \end{split}\label{int_modu}
\end{equation}
The interaction is always positive because $g(t)$ is always positive. The trap frequencies $\ome_{xy,z}$ are modulated as 
\begin{equation}
    \ome_{xy,z}(t)=\ome_{xy,z}(1+0.03\sin \ome_{\rm pot} t),
\end{equation}                                                      
where 0.03 is the modulation amplitude and $\ome_{\rm pot}$ is the modulation frequency of the potential\cite{footnote2}. This modulation amplitude is used for the entire study.

The total energy $E_{\rm tot}$ of the GP equation \eqref{GPeq} is given by
\begin{equation}
  \begin{split}
  E_{\rm tot}&=E_{\rm kin}+E_{\rm pot}+E_{\rm int}\\
  &=\int {\rm d}\bm r \bigg[\f{\hbar^2}{2m}|\nab \psi|^2+V(\bm r,t)|\psi|^2+\f{g(t)}{2}|\psi|^4\bigg],\\
  E_{\rm kin }&=\sum_{i=x,y,z} E_{{\rm k}i}=\sum_{i=x,y,z}\int {\rm d}\bm r \f{\hbar^2}{2m} |\del_i \psi|^2.
  \end{split}\label{energy_nondim}
\end{equation}
Here, $E_{\rm kin}$ is the kinetic energy, $E_{{\rm k}i}$ is the kinetic energy along the $i$ direction, $E_{\rm pot }$ is the potential energy, and $E_{\rm int}$ is the interaction energy. The decomposition of energy serves to reveal the condition of the excitation of Faraday waves and the breathing mode.

We numerically solved Eq.\eqref{GPeq} by the scaled variables $\tilde{t}=\ome_{xy}t,\,\, \bm R =(X,Y,Z)=(x,y,z)/d_{xy}$ and $\tilde{\psi}=\psi d^{3/2}_{xy}$ with $d_{xy}=\sqrt{\hbar/(m\ome_{xy})}$.
We use a pseudo-spectral method and the Runge--Kutta method of fourth order for time stepping. The grid size is $30 \times 30 \times 120$ and the number of grid points is $64\times 64\times 256$. When we modulate only the interaction, we perform simulations with the interaction $g(t)$ and constant trap frequencies. When we modulate only the potential, we perform simulations with the trap frequencies $\ome_{xy,z}(t)$ and a constant interaction.
We assume the harmonic trapping frequencies $\{\ome_{xy}/(2\pi),\ome_{z}/(2\pi)\}=\{242,27\}{\rm Hz}$ unless otherwise stated.

\section{Condition for the excitation of Faraday waves}

We investigate the condition for the excitation of Faraday waves by the modulation of the interaction of BEC.

\subsection{Typical dynamics of Faraday waves}

\begin{figure}[h]
    \centering
    \includegraphics[ width=\linewidth ]{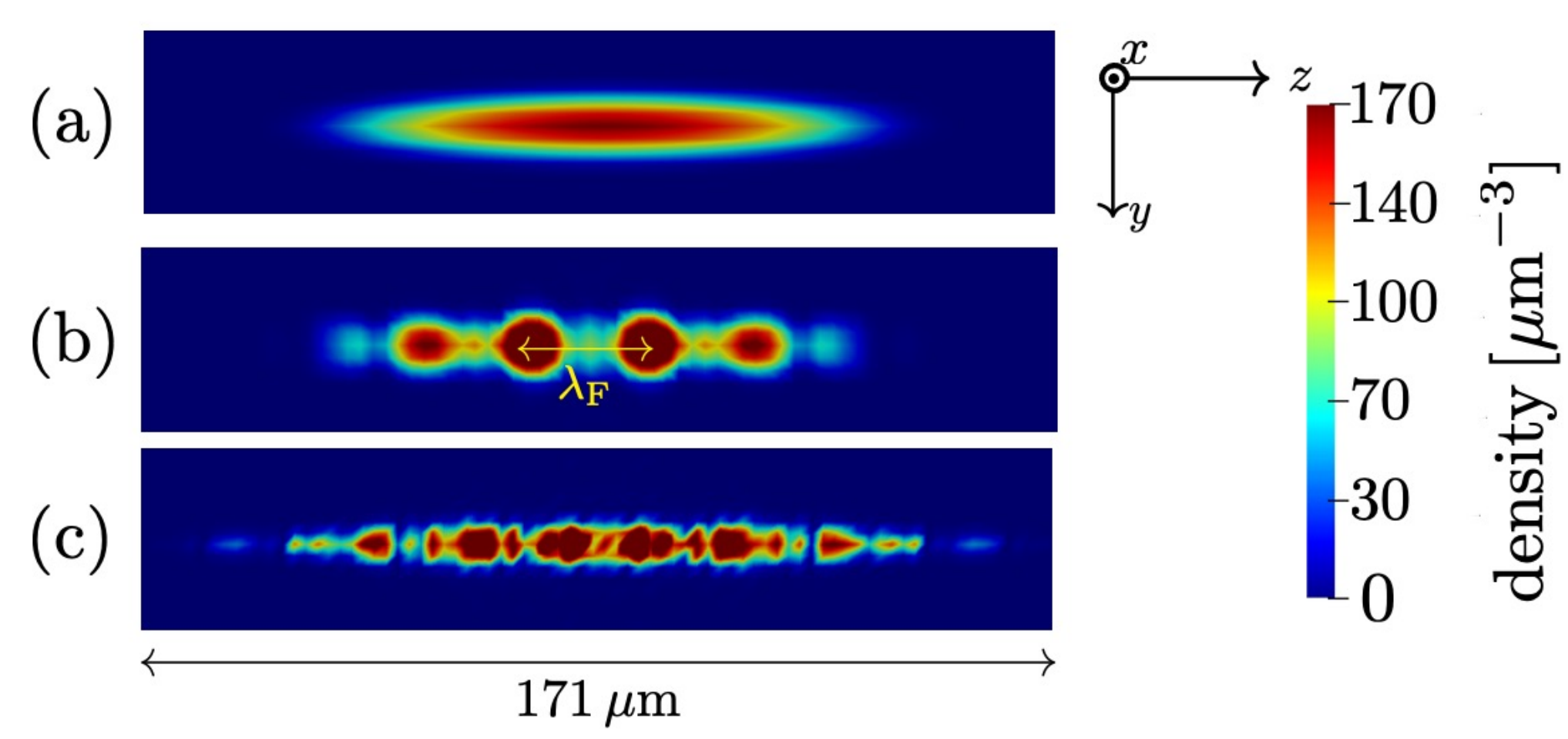}
    \caption{Temporal evolution of the nondimensional density $|\psi(\bm r,t)|^2$ in the $y$-$z$ plane. (a) Initial state at $0\,{\rm ms}$. (b) Faraday waves at $22\,{\rm ms}$. (c) Granular state at $24\,{\rm ms}$.}
    \label{fig:density_evol}
\end{figure}

We present the typical dynamics of the Faraday waves for the initial state of Fig. \ref{fig:density_evol}(a) when the interaction is modulated as
\begin{equation}
    g(t)\approx g(1+0.04\sin \ome_{\rm int} t),\,\,\,\,\ome_{\rm int}=2\ome_{xy}.
\end{equation}
The modulation frequency is resonant with the Faraday waves\cite{2}.
Figure \ref{fig:density_evol} shows the temporal evolution of the density $|\psi(\bm r,t)|^2$ in the $y$-$z$ plane. The breathing mode along the $xy$ direction is the primary excitation and that along the $z$ direction is a secondary excitation. The existence of a breathing mode along the $z$ direction was not reported in previous studies\cite{2,3,4,Mathieint,Mathiepot}. However, our numerical calculations show that it is present; the two perpendicular breathing  modes are shown to be coupled by the variational approach.
After the excitation of the breathing modes, the periodic density pattern of Faraday waves appears along the $z$ direction (Fig. \ref{fig:density_evol}(b)). Eventually, the system assumes a granular state, which has random patterns and an aperiodic density (Fig. \ref{fig:density_evol}(c)). The BEC expands along the $z$ direction with the excitation of Faraday waves and the granular state.

\begin{figure}[t]
    \centering
    \includegraphics[keepaspectratio, scale=0.4]{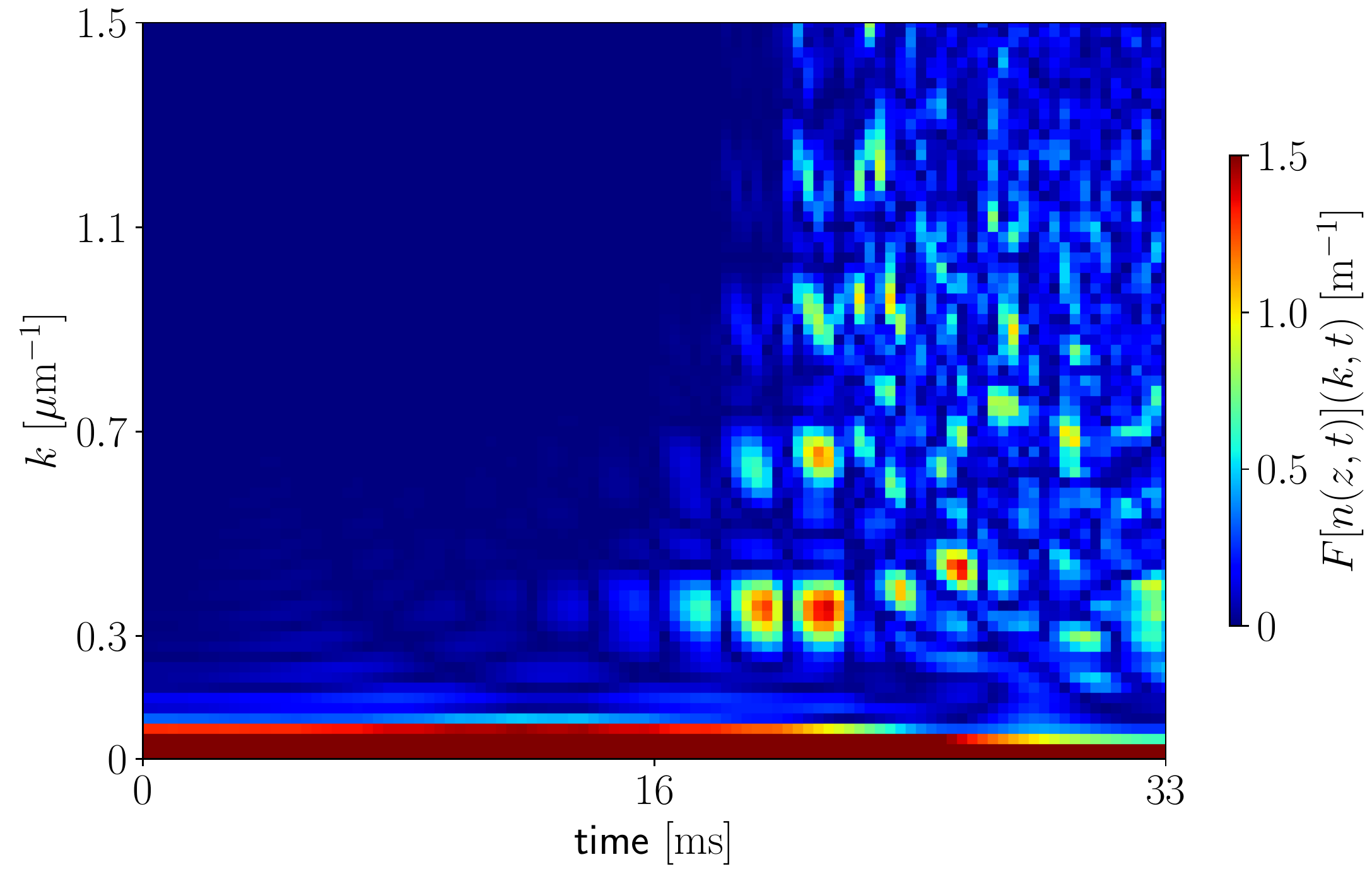}
    \caption{Time development of the Fourier transformation $F[n(z,t)](k,t)$ of $n(z,t)= \iint {\rm d}x{\rm d}y |\psi(\bm r,t)|^2$ for the dynamics of Fig. \ref{fig:density_evol}. The characteristic peaks grows as the Faraday waves appear}
    \label{fig:FFT_T9}
\end{figure}

%z方向
Figure \ref{fig:FFT_T9} shows the time development of the Fourier transformation $F[n(z,t)](k,t)$ of the density distribution $n(z,t)= \iint {\rm d}x{\rm d}y |\psi(\bm r,t)|^2$ along the $z$ direction.
The peak at $k=0$ is due to the system size of the BEC along the $z$ direction. When Faraday waves appear at $22\,{\rm ms}$, a Fourier peak appears at $k=0.3\,{\mu \rm m^{-1}}$. The Fourier peak is consistent with the spatial period $\lambda_{\rm F}\sim 20\,{\rm \mu m}$ of the Faraday waves. As the system assumes a granular state (Fig. \ref{fig:density_evol}(c)), many peaks appear corresponding to random density patterns finer than $\lambda_{\rm F}$.

\begin{figure}[t]
  \centering
  \includegraphics[width=\linewidth]{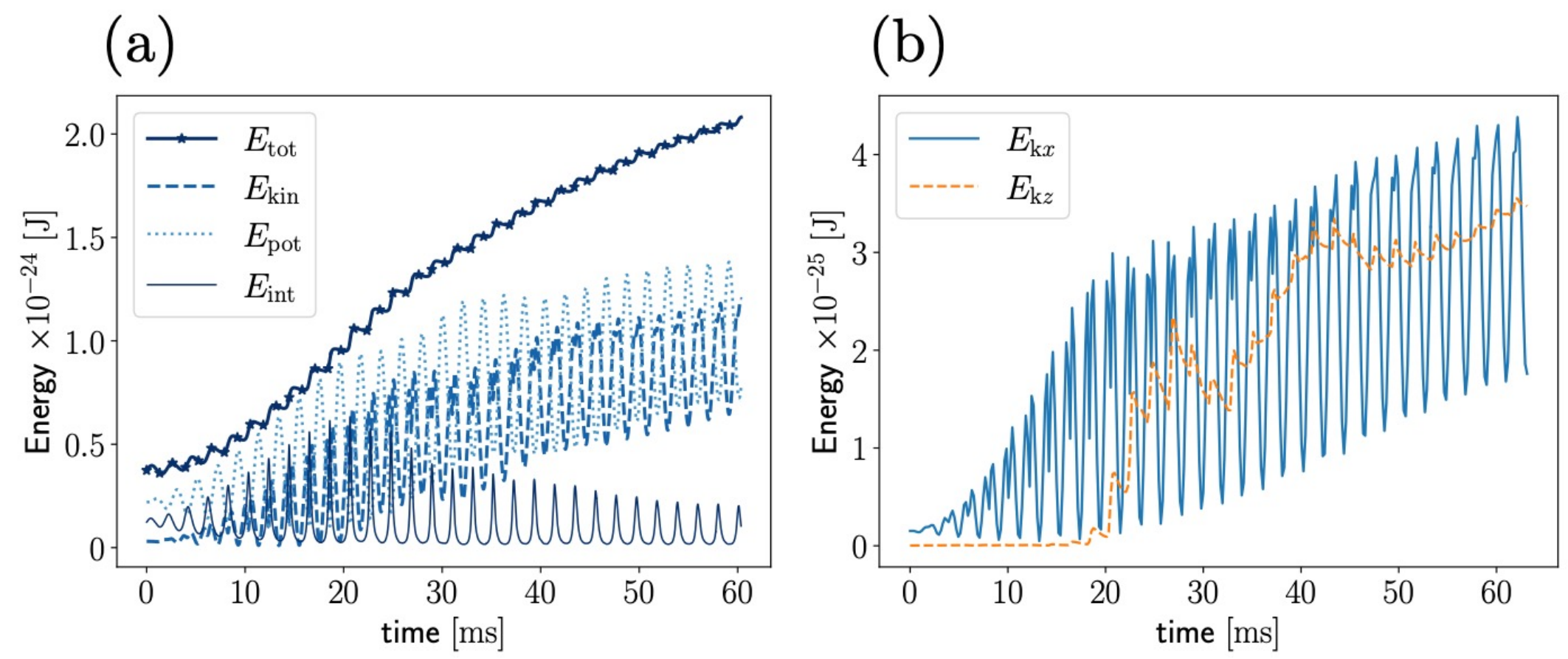}
    \caption{ Time evolution of the energy of Fig. \ref{fig:density_evol}. (a) Time development of each energy. (b) Time evolution of the decomposed kinetic energy. Each energy oscillates with the period $2\pi/(2\ome_{xy})\sim 2\,{\rm ms}$}
    \label{fig:T9_energy}
\end{figure}
Figure \ref{fig:T9_energy} shows the time evolution of the energy associated with the excitation of Faraday waves.
The energy is effectively injected to the system by the modulation of the interaction resonant with Faraday waves (Fig. \ref{fig:T9_energy}(a)).
The increase in $E_{\rm tot}$ is attributable to the increase in $E_{\rm kin}$ and $E_{\rm pot}$. 
The increase in $E_{\rm kin}$ is due to the breathing modes along the $xy$ and $z$ directions and density gradients such as Faraday waves or a granular state. These breathing modes are always present. However, the contribution of $E_{{\rm k}z}$ until $20\, {\rm ms}$ in Fig. \ref{fig:T9_energy}(b) is small because $E_{{\rm k}x,y}/E_{{\rm k}z} \sim 100$\cite{energy_ratio}. $E_{{\rm k}z}$ increases very slowly compared to $E_{{\rm k}x,y}$. 
The increase in $E_{{\rm k}z}$ after $20\,{\rm ms}$ reflects density gradients such as Faraday waves or a granular state. The increase of $E_{\rm pot}$ almost comes from the breathing mode along the $xy$ direction until $20\,{\rm ms}$. Then, BEC expands along the $z$ direction as Faraday waves are excited and BEC assumes a granular state. Thus, the expansion along the $z$ direction further increases $E_{\rm pot}$ after $20\,{\rm ms}$.

\subsection{Dependence of the dynamics on the modulation frequency}

\begin{figure}[t]
  \centering
  \includegraphics[width=\linewidth]{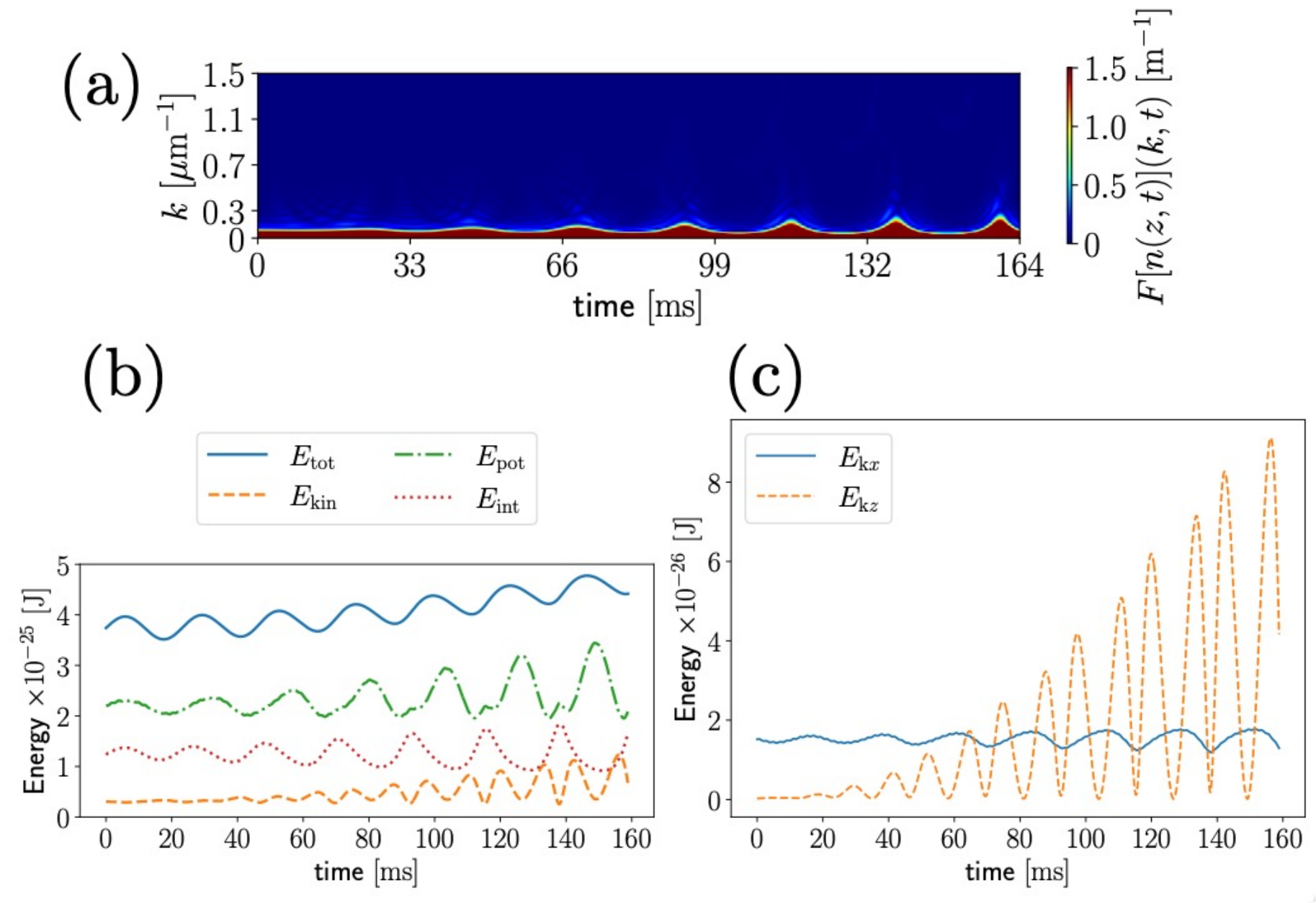}
  \caption{Time devolution of the Fourier transformation and energy when the interaction is modulated with $\ome_{\rm int} =\sqrt{3}\ome_{z}$. (a) Time evolution of $F[n(z,t)](k,t)$. (b) Time evolution of each energy. (c) Time evolution of the decomposed kinetic energy. Each energy oscillates with a period of $2\pi/(\sqrt{3} \ome_{z})\sim 20\,{\rm m s}$}
  \label{fig:BM_energy}
\end{figure}

The dynamics depends on the modulation frequency. To investigate this dependence, we modulate the system as 
\begin{equation}
    g(t)\approx g(1+0.04\sin \ome_{\rm int} t),\,\,\,\,\ome_{\rm int}=\sqrt{3}\ome_{z},
\end{equation}
with the modulation frequency resonant with the breathing mode along the $z$ direction\cite{2}. Then, the breathing mode along the $z$ direction becomes a primary excitation and that along the $xy$ direction becomes a secondary excitation.
However, Faraday waves do not appear as shown in the time development of $F[n(z,t)]$ of Fig. \ref{fig:BM_energy}(a), which is unlike the case of $\ome_{\rm int}=2\ome_{xy}$. 

The time development of each energy and decomposition of the kinetic energy is shown in Fig. \ref{fig:BM_energy}. 
Figure \ref{fig:BM_energy}(b) shows that $E_{\rm tot}$ does not follow the trend in Fig. \ref{fig:T9_energy}.
The increase of each energy comes from the system not resonant with the modulation.
Figure \ref{fig:BM_energy}(c) shows that the breathing modes along the $xy$ and $z$ directions are excited; however, the excitation along the $z$ direction becomes dominant.

\begin{figure}[t]
  \centering
  \includegraphics[width=\linewidth]{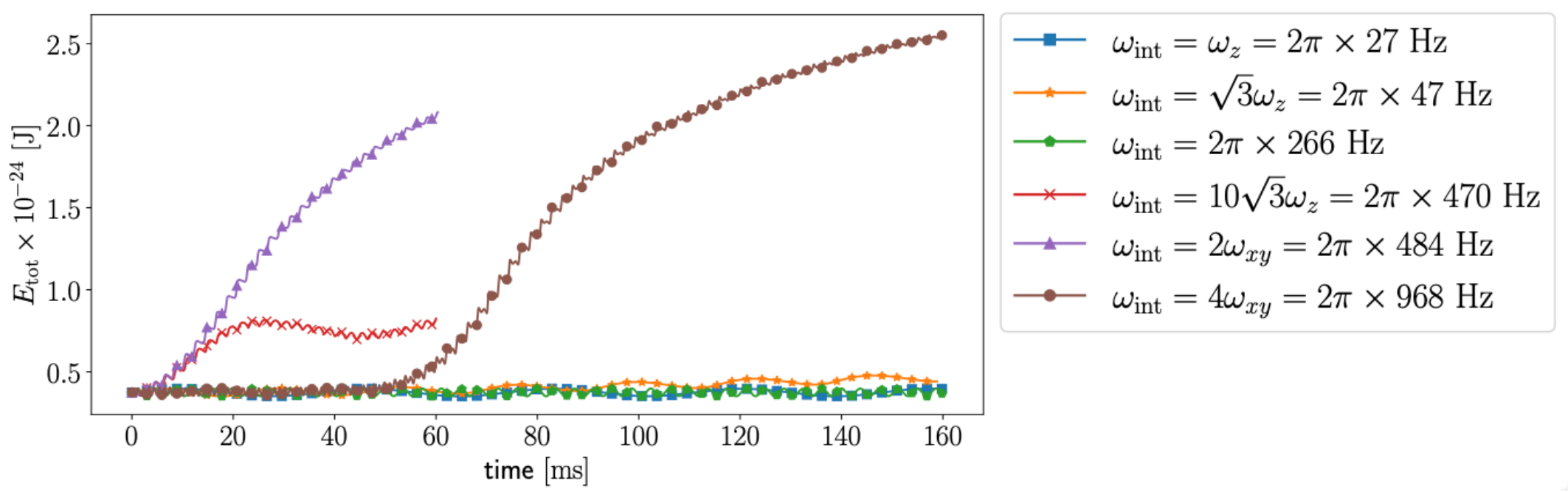}
  \caption{
   Time development of $E_{\rm tot}$ for different modulation frequencies.  When Faraday waves appear, $E_{\rm tot}$ increases rapidly }
  \label{fig:energy}
\end{figure}

We modulate the interaction using several frequencies. Figure \ref{fig:energy} shows the time dependence of $E_{\rm tot}$ on the modulation frequencies.
Faraday waves are excited for $\ome_{\rm int}=2\ome_{xy}, 10\sqrt{3}\ome_{z}\approx 2\pi\times 484=2\ome_{xy}$ following the breathing mode along the $xy$ direction. However, for $\ome_{\rm int}= \ome_{z}, \sqrt{3}\ome_{z}, 2\pi\times 266$, only the breathing mode along the $z$ direction is excited, and no Faraday waves are observed. A complicated mode is excited for $\ome_{\rm int} =4\ome_{xy}$. A detailed study of the dependence on the modulation frequencies will be conducted in the future. In the case of $\ome_{\rm int}=2\ome_{xy}$ and $10\sqrt{3}\ome_z$, the system becomes a granular state at 60 $\rm ms$. Thus, the calculation is stopped at 60 $\rm ms$, because we are interested in the excitation of Faraday waves.

\subsection{Dependence of the dynamics on the anisotropy of the potentials}

\begin{figure}[t]
  \centering
    \includegraphics[width=\linewidth]{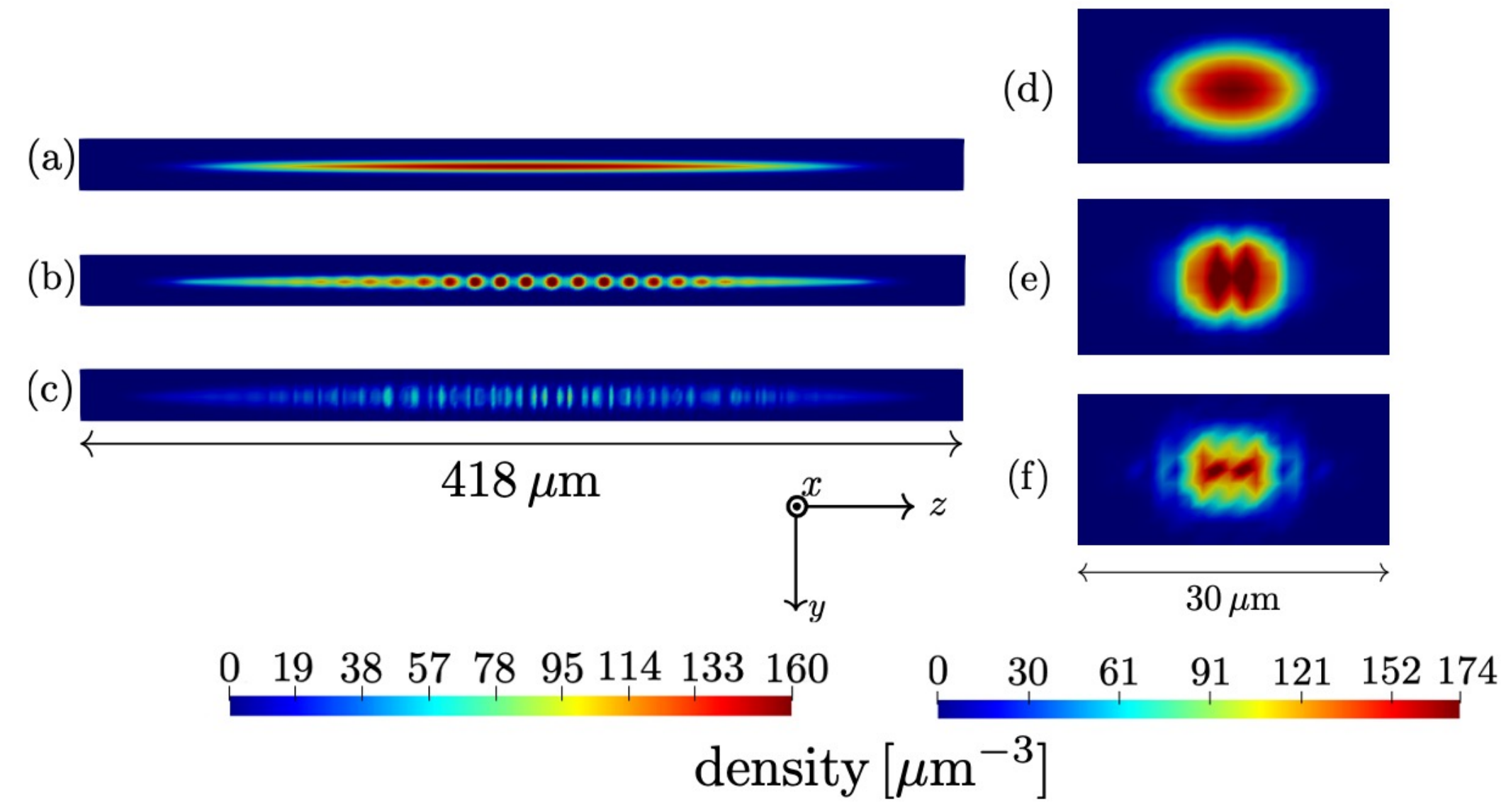}
    \caption{ Time evolution of the density in the $y$-$z$ plane. For the trap frequencies of the potential $\{\ome_{xy}/(2\pi),\ome_{z}/(2\pi)\}=\{475,7\}\,{\rm Hz}$ (a) at $0 \,{\rm ms}$, (b) at $16 \,{\rm ms}$, (c) at $19\,{\rm ms}$. For the trap frequencies of the potential $\{\ome_{xy}/(2\pi),\ome_{z}/(2\pi)\}=\{140,80\}\,{\rm Hz}$ (d) at $0\, {\rm ms}$, (e) at $28 \,{\rm ms}$, (f) at $36 \,{\rm ms}$ }
    \label{density_evol2}
\end{figure}

The anisotropy $\ome_{xy}/\ome_{z}$ of potentials is important for the emergence of Faraday waves. If we assume a spherically symmetric potential, only the breathing mode appears. 
We change the anisotropy of potentials using a fixed modulation frequency $\ome_{\rm int}=2\ome_{xy}$, which is most resonant with the Faraday waves. The mean density is maintained throughout the calculation by maintaining the volume of the BEC even with changes in geometry. Figure \ref{density_evol2} shows the two different cases of $\ome_{xy}/\ome_{z}$. In the more elongated case $\{\ome_{xy}/(2\pi),\ome_{z}/(2\pi)\}=\{475,7\}\,{\rm Hz}$ than the case of Fig. \ref{fig:density_evol}, Faraday waves are observed, as shown in Fig. \ref{density_evol2}(b).
\noindent For a more spherically symmetric potential $\{\ome_{xy}/(2\pi),\ome_{z}/(2\pi)\}=\{140,80\}\,{\rm Hz}$ than the case of Fig. \ref{fig:density_evol}, the system exhibits only a quadrupole-like mode, as shown in in Fig. \ref{density_evol2}(e).  Then, the BEC forms a granular state, as shown in Fig. \ref{density_evol2}(f). 

\begin{figure}[t]
    \centering
    \includegraphics[width=\linewidth]{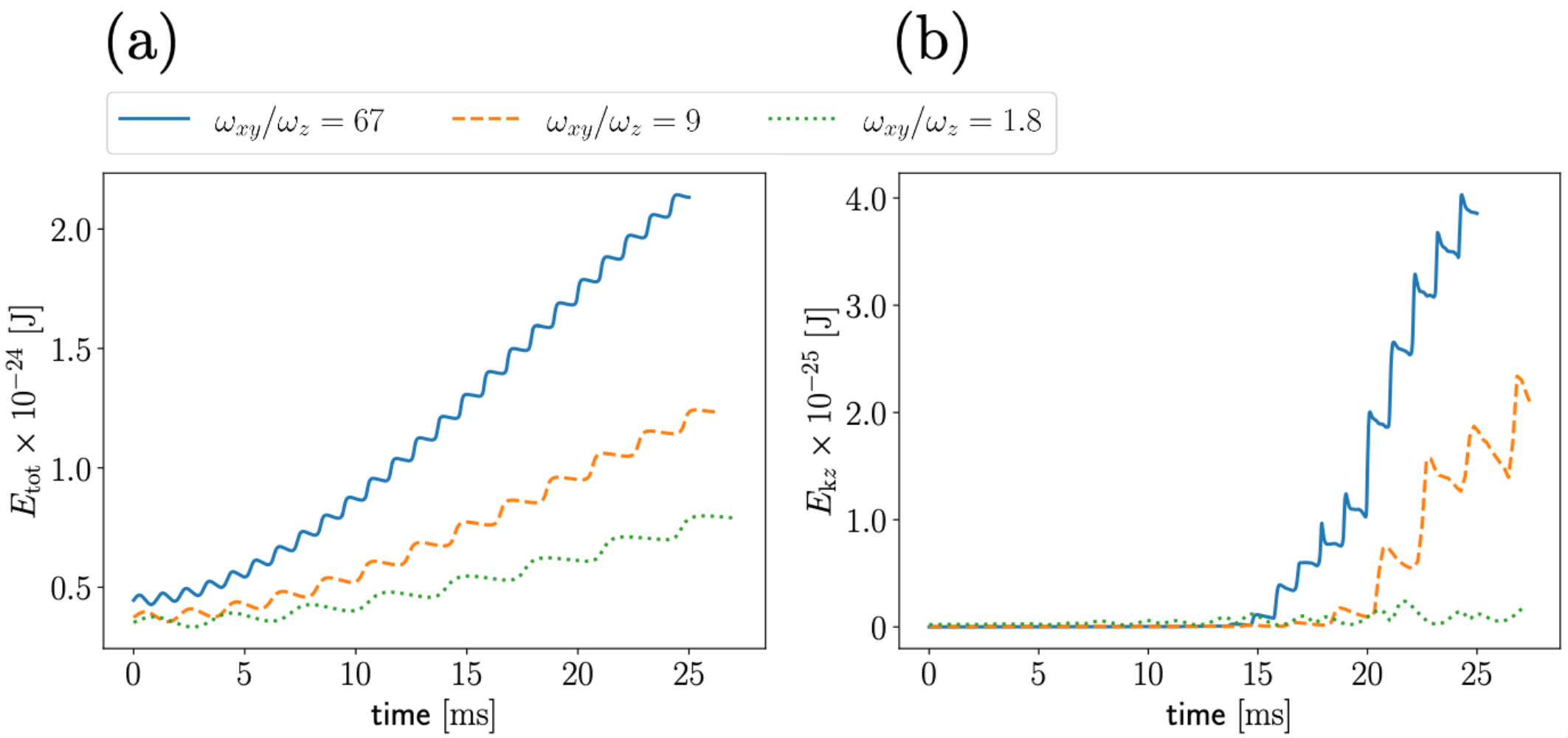}
    \caption{(a) Time evolution of $E_{\rm tot}$ for different anisotropies of potentials. (b) Time evolution of $E_{{\rm k}z}$.}
    \label{energy_eff}
\end{figure}

We find Faraday waves are more easily excited when the BEC is more elongated. Figure \ref{energy_eff} shows $E_{\rm tot }$ and $E_{{\rm k} z}$ increase faster for an elongated BEC. The time development of $E_{\rm tot}$ after 25 $\rm ms$ is not shown, because the system becomes granular state at the time for all cases.

We observe similar trends in the increase of $E_{\rm tot}$ for different anisotropies of potentials(Fig. \ref{energy_eff}(a)). To find a different signature of Faraday waves, we study the $E_{\rm kin}/E_{\rm tot}$ shown in Fig. \ref{energy_eff}(b). When the BEC is more elongated, $E_{\rm kin}/E_{\rm tot}$ increases faster, and Faraday waves are more easily excited.

\section{Comparison between different modulation methods}

Faraday waves can be excited in through interaction and potential modulation. 
We compare the differences of the dynamics of these modulation methods.

\subsection{Comparison between interaction and potential modulation}

We study potential modulation to confirm whether the dynamics are the same as for interaction modulation.

\begin{figure}[t]
    \centering
    \includegraphics[width=\linewidth]{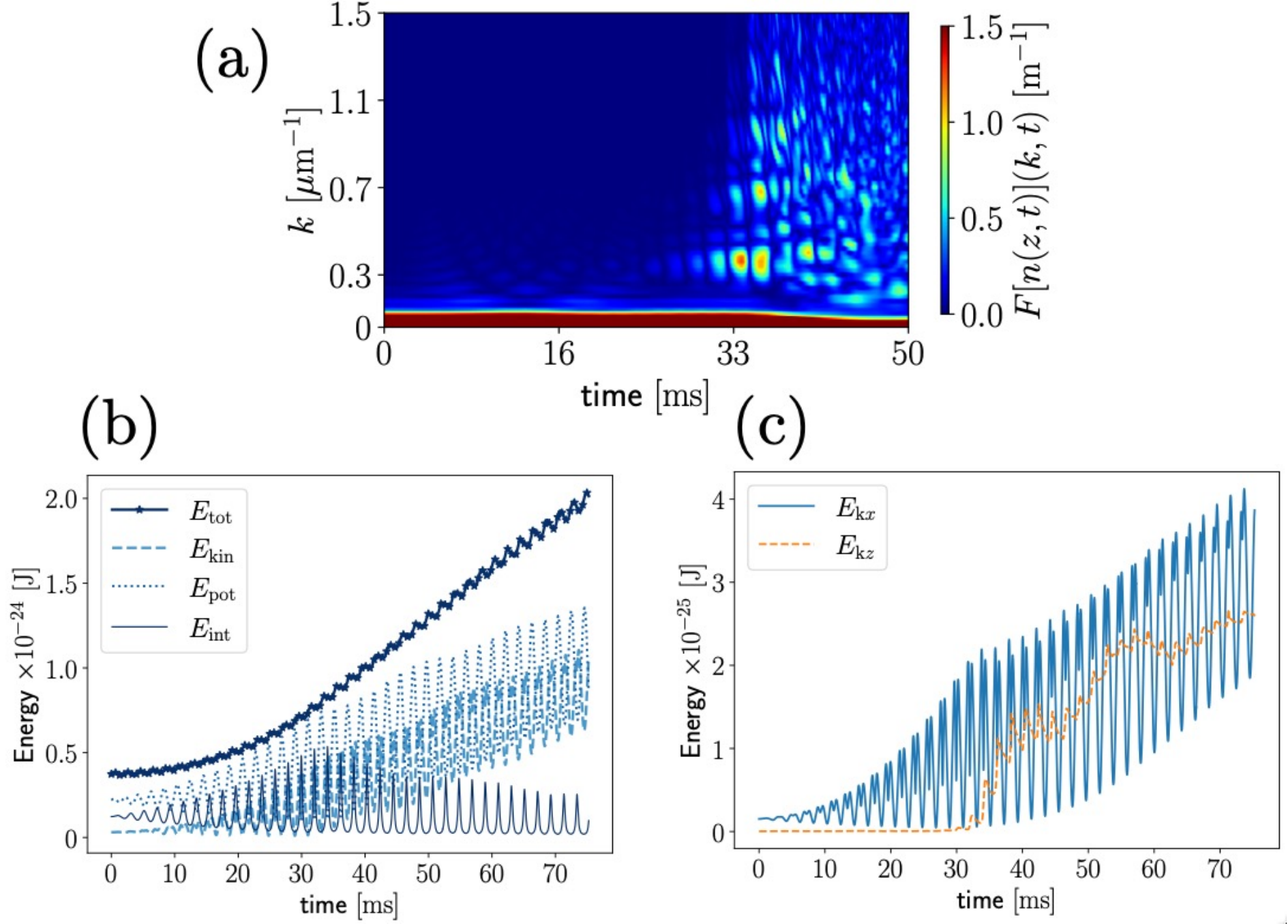}
    \caption{Time development of the Fourier transformation and time development of the energy for potential modulation that is resonant with Faraday waves. (a) Time development of $F[n(z,t)](k,t)$. (b) Time evolution of each energy. (c) Time evolution of the decomposed kinetic energy}
    \label{pot_T9_energy}
\end{figure}

Faraday waves are excited when the potential is modulated as\cite{3,4}
\begin{equation}
   \begin{split}
        V(\bm r,t)&=\f{m}{2}(\ome^2_{xy}(t)(x^2+y^2)+\ome^2_zz^2),\\
        \ome_{xy}(t)&=\ome_{xy}(1+0.03\sin \ome_{\rm pot} t),\,\,\,\,\ome_{\rm pot}=2\ome_{xy}.
   \end{split}
\end{equation}
The modulation frequency is same as in the case of interaction modulation. The excitation of Faraday waves described in 3.1 is also observed in this case, as shown in the time development of $F[n(z,t)]$ of Fig. \ref{pot_T9_energy}(a).
The time evolution of each energy and decomposition of the kinetic energy are shown in Fig. \ref{pot_T9_energy}(b), (c). There is no significant difference between Figs. \ref{fig:FFT_T9}, \ref{fig:T9_energy} and 
\ref{pot_T9_energy}.

\begin{figure}[t]
    \centering
    \includegraphics[width=\linewidth]{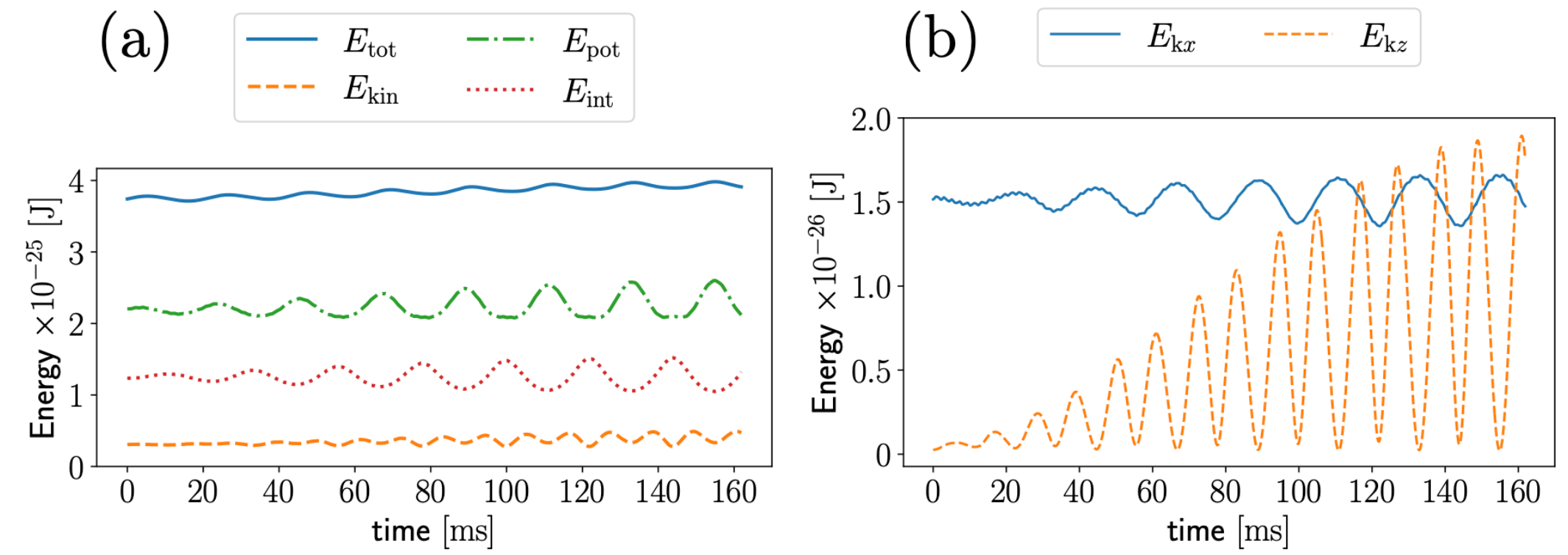}
    \caption{ Time evolution of energy for potential modulation with $\sqrt{3}\ome_z$. (a) Time evolution of each energy. (b) Time evolution of the decomposed kinetic energy}
    \label{pot_T9_BM_z_energy}
\end{figure}

Interaction modulation that is resonant with the breathing mode along the $z$ direction excited the breathing mode but not Faraday waves. 
Dynamics similar to those of potential modulation are expected and given by
\begin{equation}
    \begin{split}
        V(\bm r,t)&=\f{m}{2}(\ome^2_{xy}(x^2+y^2)+\ome^2_z(t)z^2),\\
        \ome_{z}(t)=&\ome_{z}(1+0.03\sin \ome_{\rm pot} t),\,\,\,\,\ome_{\rm pot}=\sqrt{3}\ome_z.
    \end{split}\label{pot_BM}
\end{equation}
Then, the breathing mode along the $z$ direction is excited but Faraday waves are not, which is the same as in the case of interaction modulation.
The time development of each energy and decomposition of the kinetic energy is shown in Fig. \ref{pot_T9_BM_z_energy}, which is similar to Fig. \ref{fig:BM_energy}.

Potential modulation excites Faraday waves and the breathing mode along the $z$ direction similar to interaction modulation. There is no significant difference between the two types of modulation.

\subsection{Simultaneous interaction and potential modulation }

In previous studies, Faraday waves are excited by interaction or potential modulation. We study the case where the modulations are simultaneous:
\begin{equation}
    \begin{split}
        g(t)&\approx g(1+0.04\sin \ome_{\rm int}t),\\
        V(\bm r,t)&=\f{m}{2}(\ome^2_{xy}(t)(x^2+y^2)+\ome^2_{z}(t)z^2),\\
        \ome_{xy,z}(t)&=\ome_{xy,z}(1+0.03\sin \ome_{\rm pot} t).
    \end{split}
\end{equation}
In this work, either $\ome_{xy}(t)$ or $\ome_{z}(t)$ is modulated.
We study four kinds of simultaneous modulations, as shown in Table \ref{table:modulation_methods}. The interaction and potential are modulated with the modulation frequency $2\ome_{xy}$ resonant with Faraday waves or with $\sqrt{3}\ome_z$ resonant with the breathing mode along the $z$ direction. In Table \ref{table:modulation_methods}, for modulation (i), both modulations are resonant with Faraday waves, for modulations (ii) and (iii), they are alternatively resonant with Faraday waves and the breathing mode, and for modulation (iv), they are both resonant with the breathing mode.

\begin{figure}[t]
    \centering
    \includegraphics[width=\linewidth]{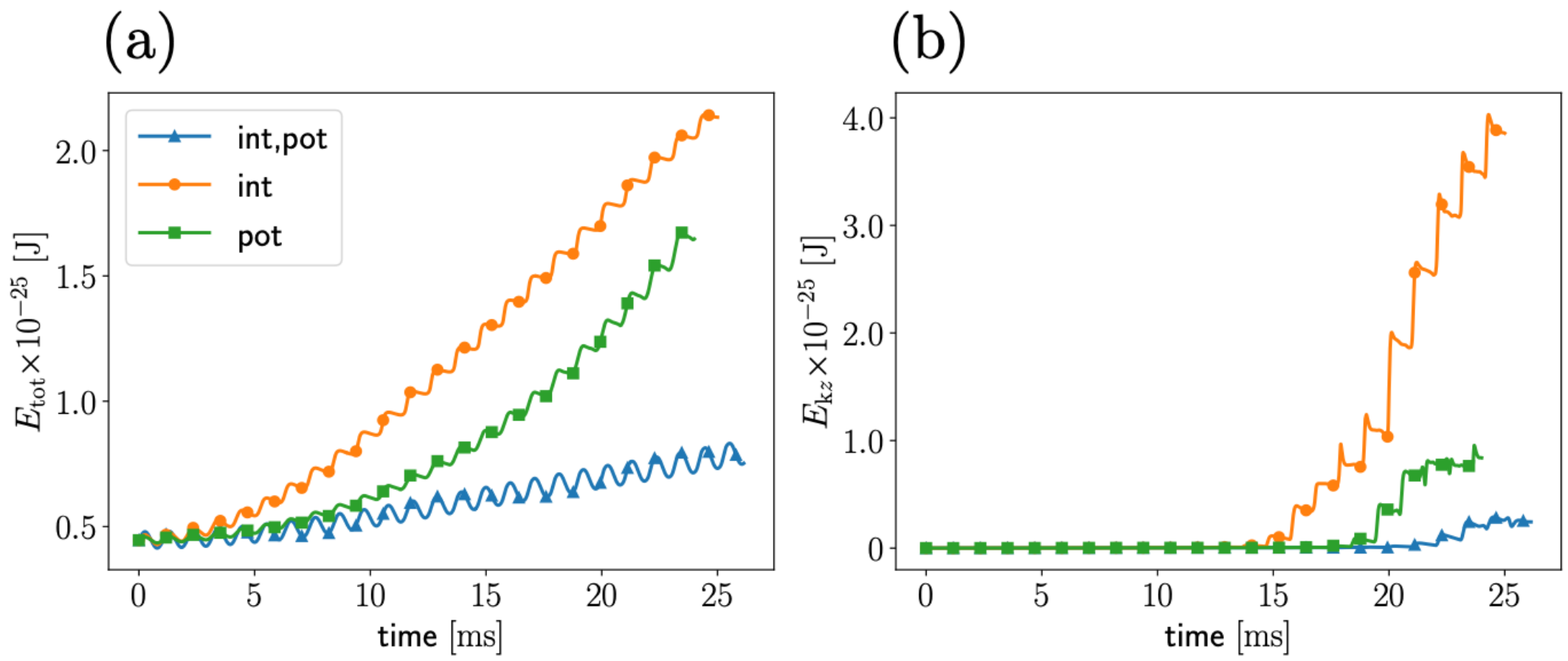}
\caption{Time evolution of the energy when the interaction and/or potential are modulated for the trap frequencies $\{\ome_{xy}/(2\pi),\ome_{z}/(2\pi)\}=\{475,7\}\,{\rm Hz}$ . (a) Time evolution of $E_{\rm tot}$. (b) Time evolution of $E_{{\rm k}z}$. In the legend, int,pot refers to simultaneous interaction and potential modulation resonant with Faraday waves. int(pot) refers to modulation of only the interaction(potential) resonant with Faraday waves}
    \label{int_pot_energy}
\end{figure}

For modulation (i), the breathing mode is excited along the $xy$ direction followed by Faraday waves along the $z$ direction. Figure \ref{int_pot_energy} compares $E_{\rm tot}$ and $E_{{\rm k}z}$ for three cases where the interaction and potential are modulated simultaneously and either of them is modulated. The case of simultaneous modulation may be  expected to be more resonant than the cases of a single type of modulation. 
However, Fig. \ref{int_pot_energy}(a) shows that $E_{\rm tot}$ for simultaneous modulation does not increase as expected. 
The onset of Faraday wave excitation for simultaneous modulation is delayed compared with the case of either modulation, as shown by the time evolution of $E_{{\rm k}z}$ in Fig. \ref{int_pot_energy}(b).

\begin{table}[t]
    \centering
    \begin{tabular}{cllll}
        \hline 
        &(i)&(ii) &(iii) &(iv)\\
        \hline 
        $\ome_{\rm int}$ & $2\ome_{xy}$ & $2\ome_{xy}$ & $\sqrt{3}\ome_{z}$&$\sqrt{3}\ome_{z}$ \\
        $\ome_{\rm pot}$ & $2\ome_{xy}$ & $\sqrt{3}\ome_{z}$ & $2\ome_{xy}$ &$\sqrt{3}\ome_{z}$\\
        \hline
    \end{tabular}
    \caption{Four kinds of modulation methods. The trap frequencies $\{\ome_{xy}/(2\pi),\ome_{z}/(2\pi)\}=\{475,7\}{\rm Hz}$ are used for (i). In (i) and (iii), only $\ome_{xy}(t)$ is modulated, while in (ii) and (iv), only $\ome_{z}(t)$ is modulated}
    \label{table:modulation_methods}
\end{table}

In modulations (ii) and (iii), Faraday waves are excited. As shown in Fig. \ref{int_pot_BM,FW}, the time evolution of $E_{{\rm tot}}$ and $E_{{\rm k}z}$ for modulation (ii) is almost same as that in Fig. \ref{fig:T9_energy} where only the interaction is modulated. The time evolution of modulation (iii) is almost the same as that of Fig. \ref{pot_T9_energy} where only the potential is modulated.
These observations are interacting. 
Even if two kinds of modulation are applied simultaneously, the excitation of Faraday waves is superior and dominates the dynamics.

\begin{figure}[t]
    \centering
    \includegraphics[width=\linewidth]{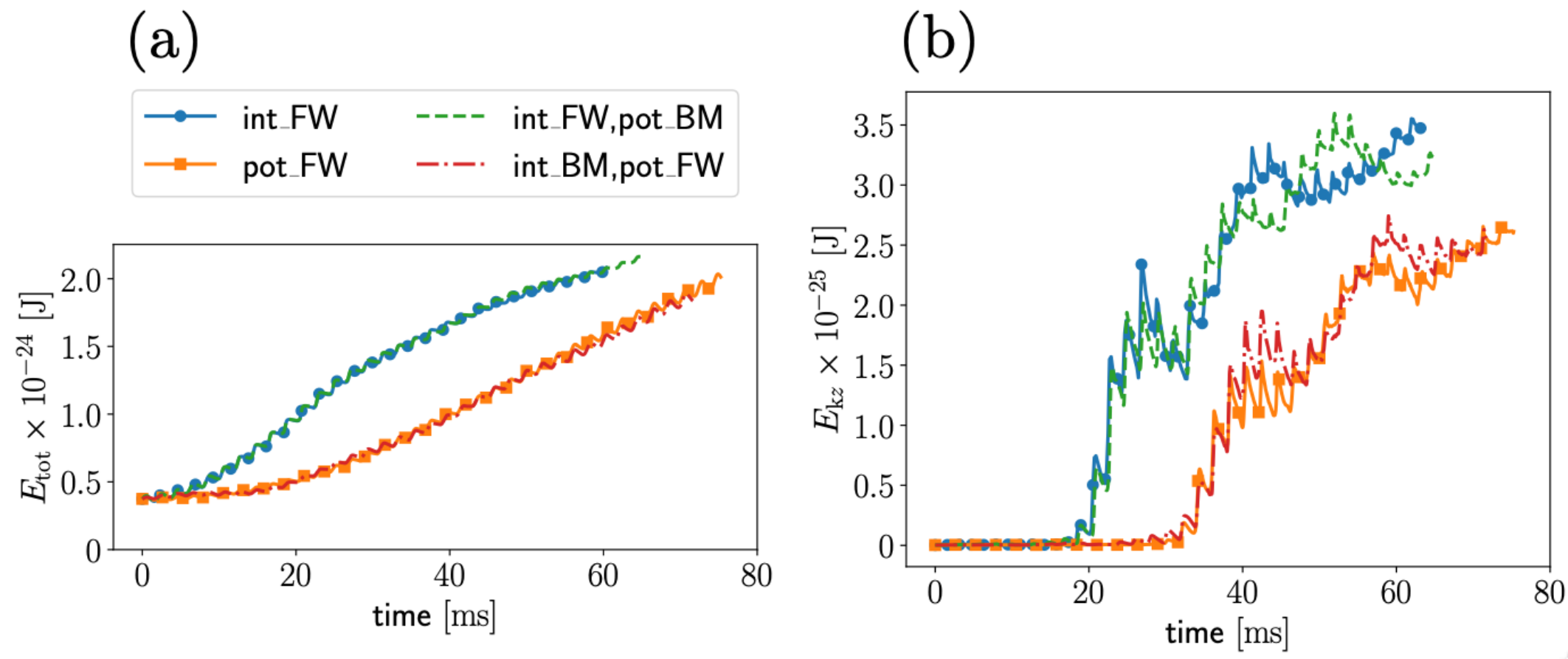}
    \caption{ Time evolution of the energy when the interaction and/or potential are modulated resonant with Faraday waves or the breathing mode along the $z$ direction. (a) Time evolution of $E_{\rm tot}$. (b) Time evolution of $E_{{\rm k}z}$. In the legend, int$\_$FW(pot$\_$FW) refers to the modulation of the interaction(potential) resonant with Faraday waves. int$\_$FW,pot$\_$BM(int$\_$BM,pot$\_$FW) refers to modulation(ii)((iii))}
    \label{int_pot_BM,FW}
\end{figure}
\begin{figure}[t]
    \includegraphics[width=\linewidth]{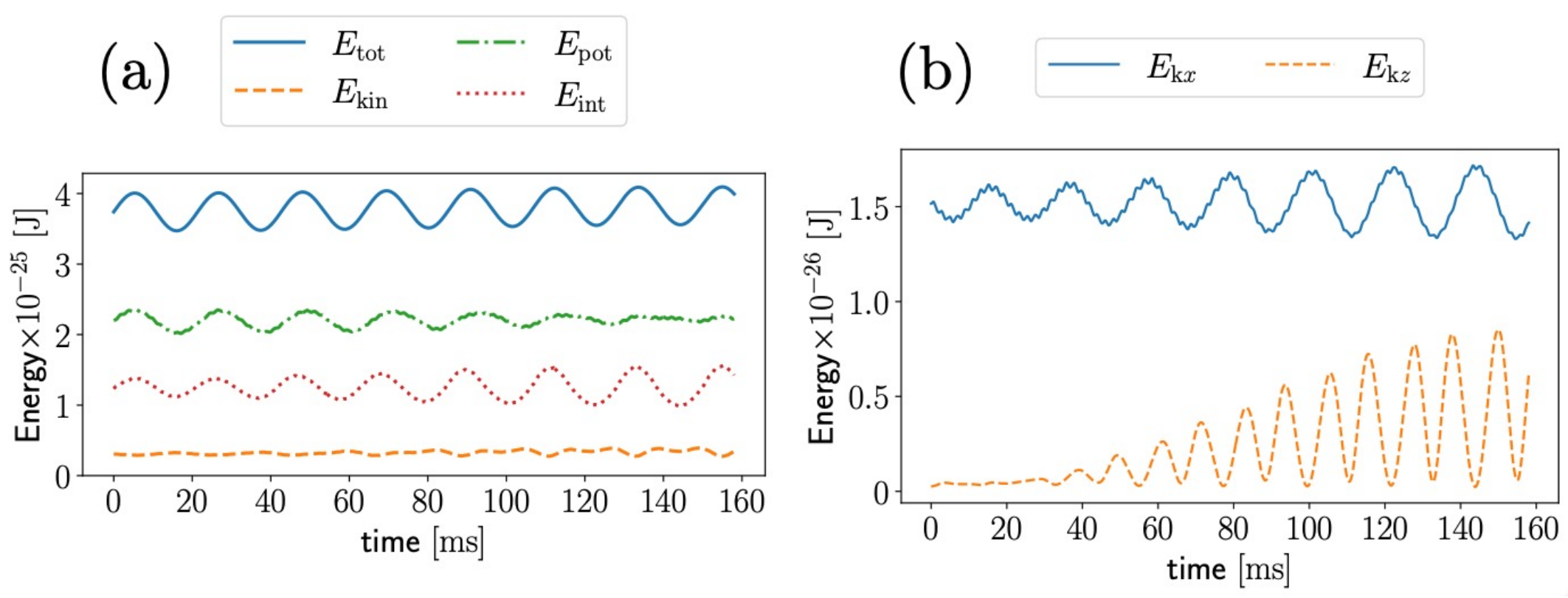}
    \caption{Time evolution of the energy when the both the interaction and potential are modulated with $\sqrt{3}\ome_z$. (a) Time evolution of each energy. (b) Time evolution of the decomposed kinetic energy}
    \label{int_pot_BM_BM_z_energy}
\end{figure}

In modulation (iv), the breathing mode along the $z$ direction is excited and Faraday waves are not excited.
However, the amplitude of $E_{{\rm k}z}$ in Fig. \ref{int_pot_BM_BM_z_energy} is smaller than that in Figs. \ref{fig:BM_energy} and \ref{pot_T9_BM_z_energy}. 
Even if two kinds of modulation are applied simultaneously, the dynamics do not resonate more strongly than for the case of either modulation. The two kinds of modulation interfere with each other. A similar phenomenon was observed for modulation (i) for the excitation of Faraday waves(Fig. \ref{int_pot_energy}).

\subsection{Comparison with theory}
We compare the spatial period $\lambda_{\rm F}$ of Faraday waves in our simulations with theory\cite{Mathiepot}, and a good agreement is observed. The dimensional spatial period $\lambda^{\rm theo}_{\rm F}(\omega)$ of Faraday waves \cite{Mathiepot} is theoretically obtained from the instability of the Mathie equation as
\begin{equation}
    \begin{split}
        \lambda^{\rm theo}_{\rm F}(\omega)&=2\pi\sqrt{\f{\pi\hbar^2\beta}{\sqrt{\al^2+\pi^2\hbar^2m^2\ome^2\beta^2}-\al}},\\
        \al&=m\rho g\omega_{xy},\,\,\,\, \beta=\sqrt{\hbar^2/m^2+\rho g/(2\pi m)},
    \end{split}
\end{equation}
where $\rho$ is the mean number of atoms per unit length along the $z$ direction.  This $\lambda^{\rm theo}_{\rm F}(\omega)$ is derived from the 3D GP model where potential is modulated and the trap frequencies are taken as $\ome_{xy}\neq 0$ and $\ome_z=0$. These situations are different from our situation of $\ome_{xy}\neq 0$ and $\ome_{z}\neq 0$. In our simulations, the spatial period $\lambda^{\rm simu}_{\rm F}(2\ome_{xy})$ of Faraday waves is obtained from the first Fourier peak of density, as in Fig. \ref{fig:FFT_T9}. $\lambda^{\rm simu}_{\rm F}(2\ome_{xy})$ is almost the same in all simulations independent of the modulation method. 
$\lambda^{\rm simu}_{\rm F}(2\ome_{xy})$ approximately agrees with $\lambda^{\rm theo}_{\rm F}(2\ome_{xy})$, as shown in Table \ref{table:lambdaf}.

\begin{table}[t]
    \centering
    \begin{tabular}{cll}
        \hline
        $\ome_{xy}/2\pi\,[{\rm Hz}]$  & $\lambda^{\rm theo}_{\rm F}\,[{\rm \mu m}] $ & $\lambda^{\rm simu}_{\rm F}\,[{\rm \mu m}]$ \\
        \hline 
        242 & 16 & 20 \\
        475 & 12 & 13 \\
        \hline
    \end{tabular}
    \caption{Comparison of the spatial period of Faraday wave simulations and theory}
    \label{table:lambdaf}
\end{table}

\section{Dynamical variables of the excitation of Faraday waves}

It would be interesting to choose a few dynamical variables characterizing the dynamical system. We propose a set of $(E_{\rm kin},E_{\rm pot},E_{\rm int})$ with $E_{\rm tot}=E_{\rm kin}+E_{\rm pot}+E_{\rm int}$. When interaction and/or potential modulation is applied to the system, we follow the trajectory of $(E_{\rm kin},E_{\rm pot},E_{\rm int})$ with the trap frequencies $\{\ome_{xy}/(2\pi),\ome_{z}/(2\pi)\}=\{242,27\}{\rm Hz}$. 

\begin{figure}[t]
    \centering
    \includegraphics[width=\linewidth]{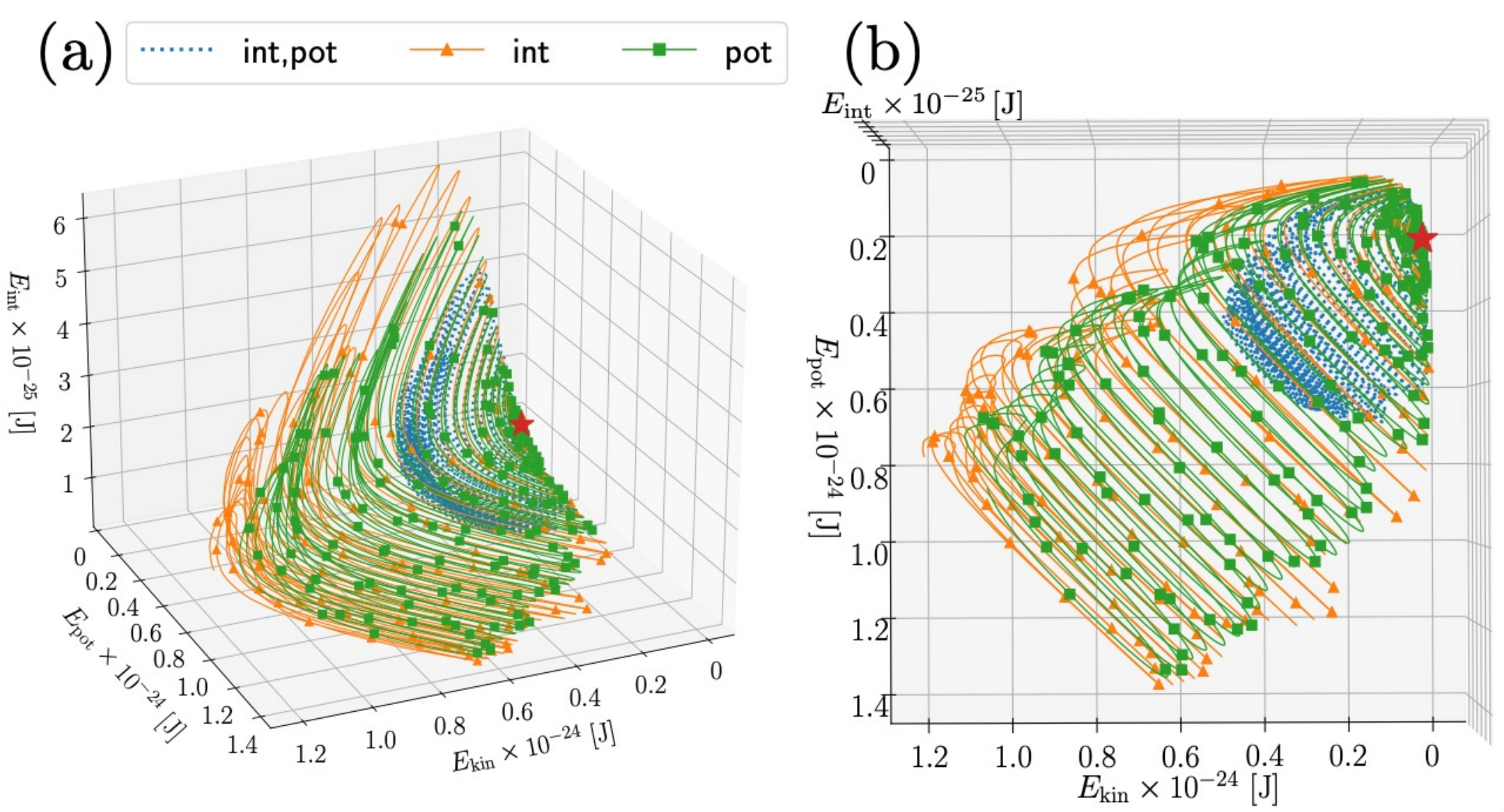}
    \caption{ (a) Trajectories in the phase space $(E_{\rm kin},E_{\rm pot},E_{\rm int})$ when Faraday waves are excited. (b) Top view of (a). The star refers to the initial state}
    \label{energy_3d_FW}
\end{figure}

Figure \ref{energy_3d_FW} shows the trajectories when Faraday waves are excited. The modulation frequency is $2\ome_{xy}$. We excite the system by three different types of modulation: modulating the interaction and potential simultaneously and modulating the interaction and potential separately. In each case, $E_{\rm kin}$ and $E_{\rm pot}$ increase with oscillating, and the trajectory always moves from the initial point in the phase space. Considering that $E_{\rm int}$ does not increase as much as $E_{\rm kin}$ and $E_{\rm pot}$, we show the top view in Fig. \ref{energy_3d_FW}(b). The trajectories retain some semblance of periodic motion and do not become chaotic. The three trajectories appear to follow the same surface despite their different kinds of modulation. 

\begin{figure}[t]
    \centering
    \includegraphics[width=\linewidth]{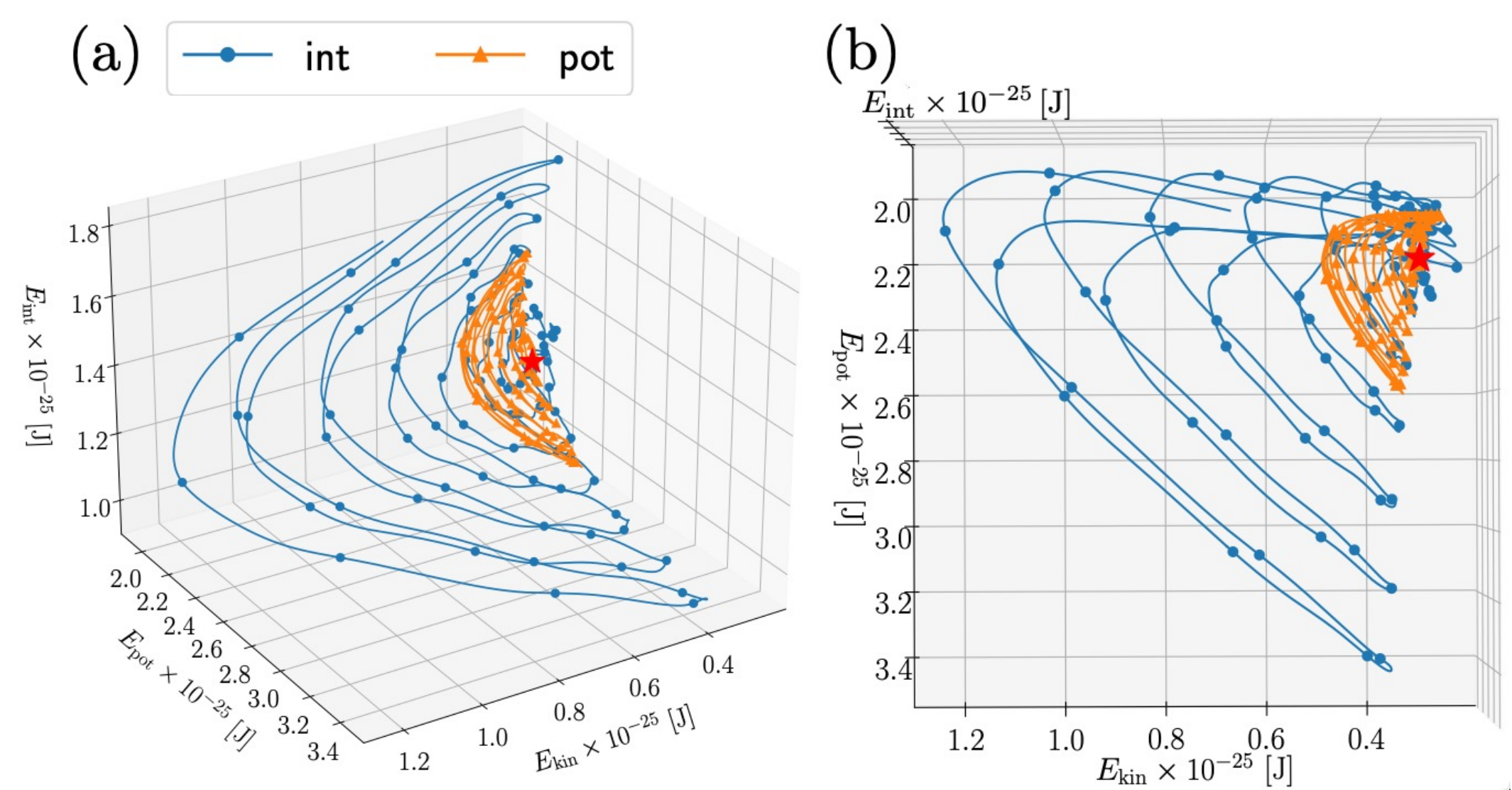}
    \caption{ (a) Trajectories in the phase space $(E_{\rm kin},E_{\rm pot},E_{\rm int})$ when the breathing mode along the $z$ direction is excited. (b) Top view of (a). The star refers to the initial state}
    \label{energy_3d_BM}
\end{figure}

The trajectories when only the breathing mode along the $z$ direction is excited are shown in Fig. \ref{energy_3d_BM}. The modulation frequency is $\sqrt{3}\ome_z$. Since each energy does not increase significantly, the trajectories remain around the initial point. Note that the scale of Fig. \ref{energy_3d_BM} is smaller than that of Fig. \ref{energy_3d_FW}.

The above analyses indicate that the dynamics are not determined by the modulation methods but the collective mode with which the modulation frequency is resonant.


\section{Conclusion}
We numerically investigated the response of an anisotropic BEC for interaction and/or potential modulation. The story of the dynamics from the breathing mode to Faraday waves is revealed by the profile of the density, its Fourier transformation and  the decomposition of energy. The excitation of Faraday waves depends on the modulation frequency and the anisotropy of potential.

In the previous studies, Faraday waves were excited by
periodic modulation of either the interaction or potential.
Studying systematically the excitations
by the two methods, we found that the kinds of the modulations are not relevant. The breathing mode along the $xy$ direction is required to excite Faraday waves.

Our study should encourage experimentalists to challenge the following problems. It is important to investigate the condition for the excitation of Faraday waves by changing the modulation frequency and the anisotropy of the potential. For example, Faraday waves are excited in elongated potential but not in spherical potential. It would be interesting to observe experimentally the critical value of $\ome_{xy}/\ome_z$ between them. It is challenging to observe that the kinds of the modulations are not relevant.

\section*{Acknowledgements}
MT acknowledges the support from JSPS KAKENHI (Grant No. JP20H01855 and No.JP23K03305).

%On the other hand, the trap potential is expected to be important for the dynamics of Faraday waves because they are excited only for elongated $\ome_{xy}\gg\ome_z$ BEC and specific modulation frequency $\ome=2\ome_{xy}$ which depend on trap frequency.


\begin{thebibliography}{99}
\bibitem{footnote2}In the previous study of 2D simulations\cite{4}, Faraday waves were excited with the trap frequency $\ome_{xy}(t)=\ome_{xy}(1+0.15\sin \ome_{\rm pot} t)$and $\ome_{\rm pot}=2.05\ome_{xy}$. We performed the current simulation using similar parameters.
\bibitem{energy_ratio}We assume a spheroid BEC as shown in Fig. \ref{fig:density_evol}(a). Let $O$ be the center of the spheroid. Let $r_{xy}$ and $r_{z}$ be the lengths along the $xy$ and $z$ axes from $O$ to the isodensity surface, respectively. According to Fig.\ref{fig:density_evol}(a), $r_{z}\sim 10 r_{xy}$. Let $n_0$ be the density at $O$ and $n$ be the density at the isodensity surface. Then, the average density gradient along the $xy$ direction is the order of $(n_0-n)/r_{xy}$ and that along the $z$ direction is the order of $(n_0-n)/r_{z}\sim (n_0-n)/10r_{xy}$. Thus, since $E_{{\rm k}x,y,z}$ are proportional to square of the absolute value of the density gradient, $E_{{\rm k}x,y}/E_{{\rm k} z}\sim 100$.
\end{thebibliography}
\end{document}